\documentclass[twocolumn,showpacs,preprintnumbers,amsmath,amssymb]{revtex4}

\begin{document}

\title{Magnetic Monopoles in Ferromagnetic Spin-Triplet Superconductors}

\author{Li-Da Zhang\footnote{Corresponding author.
                       Email: zhangld04@lzu.cn},
        Yi-Shi Duan,
        Yu-Xiao Liu}
\address{Institute of Theoretical Physics, Lanzhou University,\\
   Lanzhou 730000, P. R. China}

\begin{abstract}
Using the $\phi$-mapping method, we argue that ferromagnetic
spin-triplet superconductors allow formation of unstable magnetic
monopoles. In particular, we show that the limit points and the
bifurcation points of the $\phi$-mapping will serve as the
interaction points of these magnetic monopoles.
\end{abstract}



\maketitle

Topological solitons play important roles in many fields of
physical science ranging from condensed matter physics to QCD
\cite{Manton2004}. In Ref. \cite{BabaevPRL200288}, Babaev derived
a dual presentation of free energy for ferromagnetic spin-triplet
superconductors in terms of gauge invariant variables. The
similarity between this dual presentation and the energy function
of the Faddeev model \cite{Faddeev1975} reveals the nontrivial
topological structure of these superconductors. Based on this
topological structure, one can conclude that these superconductors
allow formation of stable knotted solitons \cite{BabaevPRL200288}.
In this paper, making use of the $\phi$-mapping method
\cite{DuanNPB1998514}, we argue that ferromagnetic spin-triplet
superconductors allow formation of unstable magnetic monopoles. In
particular, we show that at the limit points of the
$\phi$-mapping, these magnetic monopoles will be created or
annihilated in pairs, and at the bifurcation points of the
$\phi$-mapping, they will interact with each other.

First, we review the dual presentation of free energy for
ferromagnetic spin-triplet superconductors. We write the order
parameter of the spin-triplet Bose condensate as $\Psi({\bf
x},t)=\sqrt{n}({\bf x},t)\zeta({\bf x},t)$, where $n$ is the total
density and $\zeta$ is a normalized spinor. Then the free energy
of the spin-triplet superconductor reads \cite{BabaevPRL200288}
\begin{eqnarray} \label{tri}
F&=& \int d {\bf x}\Biggl[ \frac{\hbar^2}{2 M} (\nabla \sqrt{n}
)^2 + \frac{\hbar^2 n }{2 M } \left|\left(\nabla  +  i
\frac{2e}{\hbar c}{\bf A}\right)\zeta\right|^2 \nonumber \\
&-& \mu n +\frac{n^2}{2} \left[ c_0 +c_2 <{\bf F}>^2\right]
+\frac{{\bf B}^2}{8\pi} \Biggr],
\end{eqnarray}
where the average spin $<{\bf F}>=\zeta^\dag{\bf F}\zeta$. All
degenerate spinors are related to each other by gauge
transformation $e^{i\theta}$ and spin rotations ${\cal U}(\alpha,
\beta, \tau)$$=$$e^{-iF_{z}\alpha} e^{-iF_{y}\beta}
e^{-iF_{z}\tau}$, where $(\alpha, \beta, \tau)$ are the Euler
angles. Minimizing the energy with fixed particle number, the
ground state structure of $\Psi_{a}({\bf r})$ can be found
\cite{HoPRL199881}. In the ferromagnetic state where $c_{2}<0$,
the energy is minimized by $<{\bf F}>^2=1$ and the ground state
spinor and density are \cite{HoPRL199881}
\begin{eqnarray}\label{ferro}
\zeta& =& e^{i\theta} {\cal U} \left( \begin{array}{c} 1
\\0\\0\end{array} \right) = e^{i(\theta-\tau)} \left(
\begin{array}{c}
e^{-i\alpha}{\rm cos}^{2}\frac{\beta}{2} \\
\sqrt{2} {\rm cos}\frac{\beta}{2}{\rm sin}\frac{\beta}{2}
 \\ e^{i\alpha}{\rm sin}^{2}\frac{\beta}{2} \end{array} \right),
\nonumber \\
&&n^{o}({\bf r}) = \frac{1}{c_0+c_2}\mu.
\end{eqnarray}
Because the distinct configurations of $\zeta$ in Eq.
(\ref{ferro}) are given by the full range of the Euler angles, the
symmetry group of the ferromagnetic state is $SO(3)$
\cite{HoPRL199881}. Introducing new variables $\vec{\bf
s}=(s^1,s^2,s^3)=(\sin\beta \cos\alpha, \sin\beta\sin\alpha,
\cos\beta )$ and ${\bf C}=\frac{M}{en}{\bf J}$, where ${\bf
J}=\frac{i \hbar e n}{M}\left( \zeta^\dag \nabla
\zeta-\nabla\zeta^\dag\zeta\right)-\frac{4 e^2 n}{M c }{\bf A}$ is
the supercurrent, the free energy (\ref{tri}) in the ferromagnetic
state can be expressed as \cite{BabaevPRL200288}
\begin{eqnarray}\label{o3}
F&=&\int d {\bf x}\Biggl[ \frac{\hbar^2}{2 M} (\nabla \sqrt{n} )^2
+\frac{\hbar^2 n}{4 M} (\nabla \vec{\bf s})^2 + \frac{n}{8M} {\bf C}^2 \nonumber \\
&+&\frac{\hbar^2c^2}{128\pi e^2}\left(\epsilon_{abc}s_a \nabla s_b
\times \nabla s_c - \frac{1}{\hbar}\nabla\times
{\bf C}\right)^2 \nonumber \\
&-& \mu n +\frac{n^2}{2} \left[ c_0 +c_2 \right]\biggr].
\end{eqnarray}
Here and thereafter, summations over the repeated indices are
assumed. According to free energy density (\ref{o3}), the magnetic
field in the ferromagnetic spin-triplet superconductor is
separated into two part: the contribution from the supercurrent
${\bf J}$ and the self-induced magnetic field
$\tilde{\bf{B}}\equiv\frac{1}{4e} \epsilon_{abc}s_a \nabla s_b
\times \nabla s_c$, which is originated from the nontrivial
electromagnetic interaction between the components of $\zeta$.

In Ref. \cite{BabaevPRL200288}, Babaev only considered the
superconductor in a simply-connected space, which means the
defects in the superconductor do not feature the zeroes of the
order parameter. Now, using the $\phi$-mapping method, we will
investigate the case with $n=0$ at some isolated points, and find
that these points correspond to the magnetic monopoles.
Furthermore, we will study possible world-line configurations of
monopole-antimonopole pairs and multimonopoles. Since
$\pi_2(SO(3))=0$, these monopoles are topological unstable.
Technically, they are the saddle points of free energy
(\ref{tri}), and can deform into the spin textures in principle.
However, if we adjust the parameters in free energy (\ref{tri}) to
a certain range, the relative stability of the monopoles can be
guaranteed to a large extent. For simplicity, we will restrict our
study to such a parameter range in this paper.

We begin by introducing an internal vector field
\begin{equation} \label{s}
\vec{\bf S}=(S^1,S^2,S^3)=\sqrt{n}\vec{\bf s},
\end{equation}
and define a topological current as follows:
\begin{equation} \label{j}
j^{\mu}=\frac{1}{4e}\epsilon^{\mu\nu\lambda\rho}\epsilon^{abc}
 \partial_\nu s^a \partial_\lambda s^b\partial_\rho s^c.
\end{equation}
It is easy to find $j^{\mu}$ is identically conserved. The
corresponding conserved charge density $\rho=j^{0}=\nabla\cdot
\tilde{\bf{B}}$ serves exactly as the source of $\tilde{\bf{B}}$.
So we can speculate that $j^{\mu}$ is the current of the magnetic
monopole. Using $\partial_\mu s^a= \partial_\mu S^a/|\vec{\bf
S}|+S^a\partial_\mu(1/|\vec{\bf S}|)$, we can get
\begin{equation} \label{j1}
j^{\mu}=-\frac{1}{4e}\epsilon^{\mu\nu\lambda\rho}\epsilon^{abc}
 \partial_d\partial_a\left(\frac{1}{|\vec{\bf S}|}\right)\partial_\nu S^d
 \partial_\lambda S^b\partial_\rho S^c,
\end{equation}
where $\partial_a=\frac{\partial}{\partial S^a}$. Introducing the
Jacobian vector
\begin{equation} \label{d}
D^\mu\left(\frac{S}{x}\right)=\frac{1}{3!}\epsilon^{\mu \nu
\lambda\rho}\epsilon^{a b c}
\partial_{\nu}S^a \partial_{\lambda}S^b \partial _{\rho}S^c,
\end{equation}
and making use of the Green function relation in $\vec{\bf
S}$-space:
\begin{equation} \label{green}
\partial_a\partial_a\left(\frac{1}{|\vec{\bf S}|}\right)
=-4\pi \delta(\vec{\bf S}),
\end{equation}
we arrive at the following compact expression:
\begin{equation} \label{j2}
j^{\mu}=\frac{2\pi}{e}\delta(\vec{\bf
S})D^\mu\left(\frac{S}{x}\right).
\end{equation}
The $\delta$-function included in Eq. (\ref{j2}) implies that
$j^{\mu}$ can be nonzero only if $\vec{\bf S}=0$. So the zero
points of $\vec{\bf S}$ are important to determine the nontrivial
$j^\mu$. We assume $\vec{\bf S}$ has $N$ isolated zero points
denoted by ${\bf z}_r~(r=1,\cdots,N)$. According to the implicit
function theorem \cite {Goursat1904}, $\vec{\bf S}=0$ has an
unique continuous solution under the regular condition
\begin{equation} \label{regular}
D^0\left(\frac{S}{x}\right)\bigg|_{({\bf z}_r,t)}\neq0.
\end{equation}
This solution can be expressed as
\begin{equation} \label{z}
{\bf x}_r={\bf z}_r(t),
\end{equation}
which represents $N$ world lines of the magnetic monopoles. To
further illustrate the topological and physical meaning of
$j^\mu$, we need a more detailed expression for $\delta(\vec{\bf
S})$. In $\delta$-function theory \cite{Schouten1951}, given the
regular condition (\ref{regular}), we can expand $\delta(\vec{\bf
S})$ as follows:
\begin{equation}\label{delta}
\delta(\vec{\bf S})=\sum_{r=1}^N\frac{W_{r}}{D^0(\frac{S}{x})
|_{({\bf z}_r,t)}} \delta({\bf x}-{\bf z}_r(t)),
\end{equation}
where $W_{r}$ is the winding number of the $\phi$-mapping. From
the definitions of the Jacobian vector, we can obtain the velocity
vector of the magnetic monopoles:
\begin{equation}\label{velocity}
 \frac{dz_r^i(t)}{dt}=\frac{D^{i}
 (S/x)}{D^0(S/x)}\bigg|_{({\bf z}_r(t),t)}.
\end{equation}
From Eqs. (\ref{j2}), (\ref{delta}) and (\ref{velocity}), we find
\begin{eqnarray}
j^{i}&=&\frac{2\pi}{e}\sum_{r=1}^NW_{r}\frac{dz_r^i(t)}{dt}\delta({\bf
x}-{\bf z}_r(t)), \label{ji}\\
\rho&=&j^{0}=\frac{2\pi}{e}\sum_{r=1}^NW_{r}\delta({\bf x}-{\bf
z}_r(t)). \label{j0}
\end{eqnarray}
From Eqs. (\ref{ji}) and (\ref{j0}), we can see that $j^{\mu}$ is
indeed the current of the magnetic monopole. The nonvanishing of
$j^\mu$ indicates the existence of the magnetic monopole. The
corresponding magnetic charge of the $r$-th monopole is given by
the topological charge $\frac{2\pi}{e}W_{r}$. To make the energy
finite in an infinite volume ferromagnetic spin-triplet
superconductor, the magnetic monopoles can exist only in the form
of the monopole-antimonopole pairs. In such a pair, the monopole
and antimonopole will be connected by a Dirac string, or a
doubly-quantized vortex, which belongs to the trivial topological
class of $\pi_1(SO(3))=Z_2$.

From the perspective of mathematics, we describe the monopoles by
the vector field $\vec{\bf S}$, which does not include the Euler
angle $\tau$. This description can not indicate the lack of
stability of the monopoles. A more detailed description must
involve $\tau$, and hence reveals more detailed properties of the
monopoles. We leave this subject to future studies.

When condition (\ref{regular}) fails at some fixed spacetime
points $({\bf z}_{r0},t_0)$, i.e.
\begin{equation}\label{D0}
D^0\left(\frac{S}{x}\right)\bigg|_{({\bf z}_{r0},t_0)}=0,
\end{equation}
we call $({\bf z}_{r0},t_0)$ a branch point of the $\phi$-mapping,
or a branch point for short. In other words, the branch point is
the point determined by $\vec{\bf S}=0$ and $D^0(\frac{S}{x})= 0$.
There are two kinds of branch points, namely the limit points and
the bifurcation points. If for at least one space index $i$,
\begin{equation}\label{D1}
D^{i}(\frac{S}{x})\bigg|_{({\bf z}_{r0},t_0)}\neq0,
\end{equation}
we call $({\bf z}_{r0},t_0)$ a limit point. If for all space
indices $i$,
\begin{equation}\label{D2}
D^{i}(\frac{S}{x})\bigg|_{({\bf z}_{r0},t_0)}=0,
\end{equation}
we call $({\bf z}_{r0},t_0)$ a bifurcation point.

Here we first discuss the evolution processes of the magnetic
monopoles in the neighborhood of the limit point. Because of Eq.
(\ref{D0}), we cannot use the implicit function theorem at the
limit point as the above. But we can use $D^{1}(\frac{S}{x})$
instead of $D^0(\frac{S}{x})$ to continue the discussion if we
choose $i=1$ in Eq. (\ref{D1}). This means that at the limit point
$({\bf z}_{r0},t_0)$, $\vec{\bf S}=0$ has an unique continuous
solution, which can be expressed as
\begin{eqnarray}
 t&=&t(x^1), \label{t}\\
 x^{2}&=&x^{2}(x^1), \label{x2}\\
 x^{3}&=&x^{3}(x^1). \label{x3}
\end{eqnarray}
Similar to Eq. (\ref{velocity}), here we can find
\begin{equation}\label{t/z}
 \frac{dt}{dx^1}\bigg|_{z_{r0}^1}=\frac{D^0(S/x)}
 {D^1(S/x)}\bigg|_{({\bf z}_{r0},t_0)}=0.
\end{equation}
Then the Taylor expansion of Eq. (\ref{t}) in the neighborhood of
the limit point $({\bf z}_{r0},t_0)$ reads
\begin{equation}\label{tay}
t-t_0=\left.\frac
12\frac{d^2t}{(dx^1)^2}\right|_{{z_{r0}^1}}(x^1-z_{r0}^1)^2
+(\mathrm{higher~order~terms}).
\end{equation}
Ignoring the higher order terms, Eq. (\ref{tay}) represents a
parabola in the $x^1-t$ plane. If $\frac{d^2t}{(dx^1)^2}|_{({\bf
z}_{r0},t_0)}>0~(<0)$, this parabola implies that at the limit
point $({\bf z}_{r0},t_0)$, there is a monopole-antimonopole pair
created (annihilated). So we can conclude that the limit points of
the $\phi$-mapping are the points where the monopole-antimonopole
pairs are created or annihilated.

Now we turn to the evolution processes of the magnetic monopoles
in the neighborhood of the bifurcation point. Due to Eqs.
(\ref{D0}) and (\ref{D2}), the velocity vector of the magnetic
monopole at the bifurcation point $({\bf z}_{r0},t_0)$ can not be
determined by Eq. (\ref{velocity}). In general, there will be more
than one velocity vector at the bifurcation point. So, in order to
find out all velocity vectors at the bifurcation point $({\bf
z}_{r0},t_0)$, we assume that
\begin{equation}\label{subD}
\left(\frac{\partial S^1}{\partial x^2}\frac{\partial
S^2}{\partial x^3} -\frac{\partial S^1}{\partial
x^3}\frac{\partial S^2}{\partial x^2}\right)\bigg|_{({\bf
z}_{r0},t_0)}\neq0.
\end{equation}
Then, from the implicit function theorem \cite {Goursat1904}, it
follows that there is an unique continuous solution to
\begin{equation}\label{s12}
S^1({\bf x},t)=0,~S^2({\bf x},t)=0.
\end{equation}
This solution can be expressed as
\begin{equation}\label{x23}
x^2=f^2(x^1,t),~~x^3=f^3(x^1,t).
\end{equation}
Substitude Eq. (\ref{x23}) into Eq. (\ref{s12}), we obtain
\begin{eqnarray}
S^1(x^1,f^2(x^1,t),f^3(x^1,t),t)\equiv0,\label{s1}\\
S^2(x^1,f^2(x^1,t),f^3(x^1,t),t)\equiv0.\label{s2}
\end{eqnarray}
From the differentiations of Eqs. (\ref{s1}) and (\ref{s2}), as
well as the Gaussian elimination method, we can find all the first
and the second partial derivatives of $f^1(x^1,t)$ and
$f^2(x^1,t)$. Substitute Eq. (\ref{x23}) into $S^3({\bf x},t)=0$,
we obtain
\begin{equation}\label{F}
F(x^1,t)\equiv S^3(x^1,f^2(x^1,t),f^3(x^1,t),t)=0.
\end{equation}
From the definition of the branch point, we have
\begin{equation}\label{F0}
F(z_{r0}^1,t_0)=0.
\end{equation}
Using the partial derivatives of $f^2$ and $f^3$, as well as the
Cramer rule, it can be proved that
\begin{equation}\label{pF}
\frac{\partial F}{\partial x^1}\bigg|_{(z_{r0}^1,t_0)}=0,
~~\frac{\partial F}{\partial t}\bigg|_{(z_{r0}^1,t_0)}=0.
\end{equation}
Then the Taylor expansion of Eq. (\ref{F}) in the neighborhood of
the bifurcation point $({\bf z}_{r0},t_0)$ reads
\begin{eqnarray}\label{tayF}
F(x^{1},t)=\frac12A(x^1-z_{r0}^1)^2+B(x^1-z_{r0}^1)(t-t_0)\nonumber\\
+\frac12C(t-t_0)^2+(\mathrm{higher~order~terms})=0,
\end{eqnarray}
where $A=\frac{\partial^{2}F}{(\partial x^1)^2}|_{(z_{r0}^1,
t_0)}$, $B=\frac{\partial^{2}F}{\partial x^1 \partial
t}|_{(z_{r0}^1, t_0)}$ and $C=\frac{\partial^{2}F}{(\partial
t)^2}|_{(z_{r0}^1,t_0)}$. From the partial derivatives of $f^2$
and $f^3$, the constants $A$, $B$ and $C$ are calculable. Dividing
Eq. (\ref{tayF}) by $(t-t_0)^2$, and taking the limit
$x^1\rightarrow z_{r0}^1$ and $t\rightarrow t_0$, Eq. (\ref{F})
gives rise to
\begin{equation}\label{ABC}
A(\frac{dx^1}{dt})^2+2B\frac{dx^1}{dt}+C=0.
\end{equation}
Similarly, we can also find
\begin{equation}\label{CBA}
C(\frac{dt}{dx^1})^2+2B\frac{dt}{dx^1}+A=0.
\end{equation}
From Eq. (\ref{ABC}) or Eq. (\ref{CBA}), we can obtain the
velocity component $\frac{dx^1}{dt}$. The other velocity
components $\frac{dx^2}{dt}$ and $\frac{dx^3}{dt}$ can be obtained
from $\frac{dx^1}{dt}$ and the partial derivatives of $f^2$ and
$f^3$. Therefore, the different velocity vectors of the magnetic
monopoles at the bifurcation point $({\bf z}_{r0},t_0)$ can be
determined completely.

According to the different values of $A$, $B$ and $C$, there are
four possible cases:

Case 1. For $A\neq0$ and $ B^2-AC>0$, we have two different
solutions to Eq. (\ref{ABC}). This case implies that two monopoles
meet and then depart from each other at the bifurcation point.

Case 2. For $A\neq0$ and $B^2-AC=0$, we have only one solution to
Eq. (\ref{ABC}). This case implies three different evolution
processes: (a) one multimonopole split into two, (b) two monopoles
merge into one, and (c) two monopoles tangentially intersect at
the bifurcation point.

Case 3. For $A=0$, $B\neq 0$ and $ C\neq 0$, the velocity
component $\frac{dx^1}{dt} = -\frac{C}{2B}$ or tend to infinity.
This case implies two different evolution processes: (a) one
multimonopole splits into three, and (b) three monopoles merge
into one at the bifurcation point.

Case 4. For $A=0$ and $C=0$, the velocity component
$\frac{dx^1}{dt} = 0$ or tend to infinity. This case also implies
two different evolution processes which are similar to Case 3.

When condition (\ref{subD}) fails, and all other $2\times2$
sub-Jacobian also vanish at the bifurcation point $({\bf
z}_{r0},t_0)$, we need to discuss the evolution processes at the
higher order bifurcation point. This discussion will be more
complicated, but the method will be similar to that we used above.
From the above analysis, we can conclude that the bifurcation
points of the $\phi$-mapping are the points where the monopoles
interact with each other.

In conclusion, using the $\phi$-mapping method, we have argued
that ferromagnetic spin-triplet superconductors allow formation of
unstable magnetic monopoles. The limit points and the bifurcation
points of the $\phi$-mapping will serve as the interaction points
of these magnetic monopoles.

This work was supported by the National Natural Science Foundation
of the People's Republic of China (No. 10475034 and No. 10705013)
and the Fundamental Research Fund for Physics and Mathematics of
Lanzhou University (No. Lzu07002).

\end{document}